\documentclass[lettersize,journal]{IEEEtran}
\usepackage{amsmath}

\usepackage{caption}
\usepackage{lipsum}
\usepackage{amsmath,amsfonts}
\usepackage{algorithmic}
\usepackage{algorithm}
\usepackage{array}
\usepackage[caption=false,font=normalsize,labelfont=sf,textfont=sf]{subfig}
\usepackage{textcomp}
\usepackage{amsmath}
\usepackage{url}
\usepackage{verbatim}
\usepackage{graphicx}
\usepackage{cite}
\usepackage{bm}
\usepackage{booktabs}
\usepackage{multirow}
\usepackage{multicol}
\usepackage{mathptmx}                  
\usepackage{times}                     
\usepackage{cuted} 
\usepackage{dblfloatfix} 

\usepackage{colortbl}  
\usepackage{xcolor}
\usepackage{array}  
\definecolor{llgray}{rgb}{0.8, 0.8, 0.8}

\definecolor{colorOrange}{rgb}{0.98, 0.945, 0.82}

\usepackage{color}

\usepackage{makecell} 
\usepackage{hyperref}

\definecolor{customblue}{RGB}{0,0,255} 
\newcommand{\ourmethod}{SIAgent}
\newcommand{\Agent}{\textit{Agent}}
\newcommand{\Agt}{\textit{Agt}}
\newcommand{\CTRL}{\textit{CTRL}}
\newcommand{\GP}{\textit{GP}}
\newcommand{\inblue}[2]{\textcolor[rgb]{0,0,1}{#2}}

\newcommand{\majorrevise}[2]{\textcolor[rgb]{0,0,0}{#2}}

\begin{document}


\title{SIAgent: Spatial Interaction Agent via LLM-powered Eye-Hand Motion Intent Understanding in VR}


\author{Zhimin Wang, Chenyu Gu, and Feng Lu, \IEEEmembership{Senior Member,~IEEE}
\thanks{Zhimin Wang, Chenyu Gu, and Feng Lu are with the State Key Laboratory of Virtual
Reality Technology and Systems, School of Computer Science and
Engineering, Beihang University, Beijing 100191, China. e-mail:
$\left\{zm.wang \ \backslash \ gucy \ \backslash \ lufeng\right\}$@buaa.edu.cn.
}
\thanks{Feng Lu is the corresponding author.}
\thanks{Manuscript received July 17, 2025.}}

\markboth{IEEE TRANSACTIONS ON VISUALIZATION AND COMPUTER GRAPHICS, July~2025}%
{Shell \MakeLowercase{\textit{et al.}}: A Sample Article Using IEEEtran.cls for IEEE Journals}

\maketitle

\begin{figure*}[htbp]
    \centering
    \vspace{-2mm}
        \includegraphics[width=0.92\linewidth]{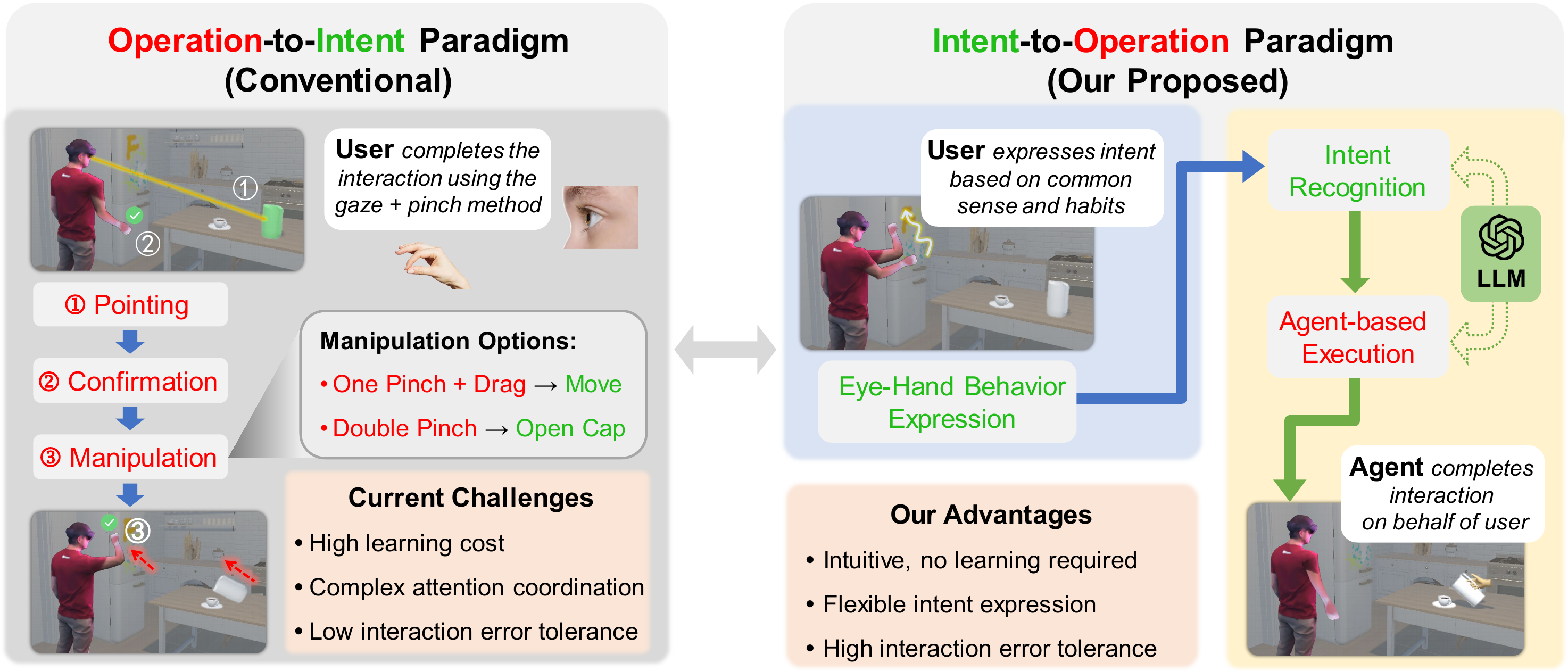}
        \caption{\majorrevise{}{The conventional Operation-to-Intent paradigm (left) requires users to perform sequential operations such as gaze-based pointing, gesture confirmation, and manipulation, resulting in high learning costs and low error tolerance. In contrast, our proposed Intent-to-Operation paradigm (right) allows users to naturally express intent through eye-hand motion, which is then recognized and executed by an agent powered by LLMs, enabling intuitive, flexible, and robust interaction.}}
        \vspace{-6mm}
        \label{fig:teaser}
\end{figure*}

\begin{abstract}

\majorrevise{}{Eye-hand coordinated interaction is becoming a mainstream interaction modality in Virtual Reality (VR) user interfaces.} Current paradigms for this multimodal interaction require users to learn predefined gestures and memorize multiple gesture-task associations, which can be summarized as an ``Operation-to-Intent" paradigm. This paradigm increases users' learning costs and has low interaction error tolerance. In this paper, we propose \textit{\textbf{\ourmethod}}, a novel ``Intent-to-Operation" framework allowing users to express interaction intents through natural eye-hand motions based on common sense and habits. Our system features two main components: (1) intent recognition that translates spatial interaction data into natural language and infers user intent, and (2) agent-based execution that generates an agent to execute corresponding tasks. This eliminates the need for gesture memorization and accommodates individual motion preferences with high error tolerance. We conduct two user studies across over 60 interaction tasks, comparing our method with two ``Operation-to-Intent" techniques. Results show our method achieves higher intent recognition accuracy than gaze + pinch interaction (97.2\% vs 93.1\%) while reducing arm fatigue and improving usability, and user preference. Another study verifies the function of eye gaze and hand motion channels in intent recognition. Our work offers valuable insights into enhancing VR interaction intelligence through intent-driven design. Our source code and LLM prompts will be made available upon publication. Demo video and supplementary materials can be found at: \href{https://zhimin-wang.github.io/SIAgent.html}{https://zhimin-wang.github.io/SIAgent.html}

\end{abstract}


\section{Introduction}

Spatial interaction between users and virtual objects is central to creating immersive experiences in Virtual Reality (VR). Current natural spatial interaction inputs for VR include hand gesture, speech, and eye gaze \cite{wang2025spatial, cao2023real, GazePointAR}. \majorrevise{}{While eye gaze offers rapid target selection, addressing the Midas Touch problem remains critical, where unintentional gaze triggers unwanted actions \cite{ 10670454}. Hand motion enables complex operations and provides silent input, yet prolonged use may induce physical fatigue \cite{DBLP:conf/vr/ChaconasH18, DBLP:conf/chi/PeiCLZ22}. Speech interaction provides a powerful hands-free modality, though its reliance on vocalization raises social acceptance concerns in shared environments \cite{DBLP:journals/tvcg/CaiML24}. To leverage these complementary strengths, eye-hand coordinated interaction combines the speed of gaze with the precision of gestures, increasingly serving as a dominant paradigm in VR.} For example, Apple Vision Pro has incorporated this multimodal interaction as a fundamental approach \cite{Apple}.

\begin{figure}
    \begin{center}
    \includegraphics[width=1\linewidth]{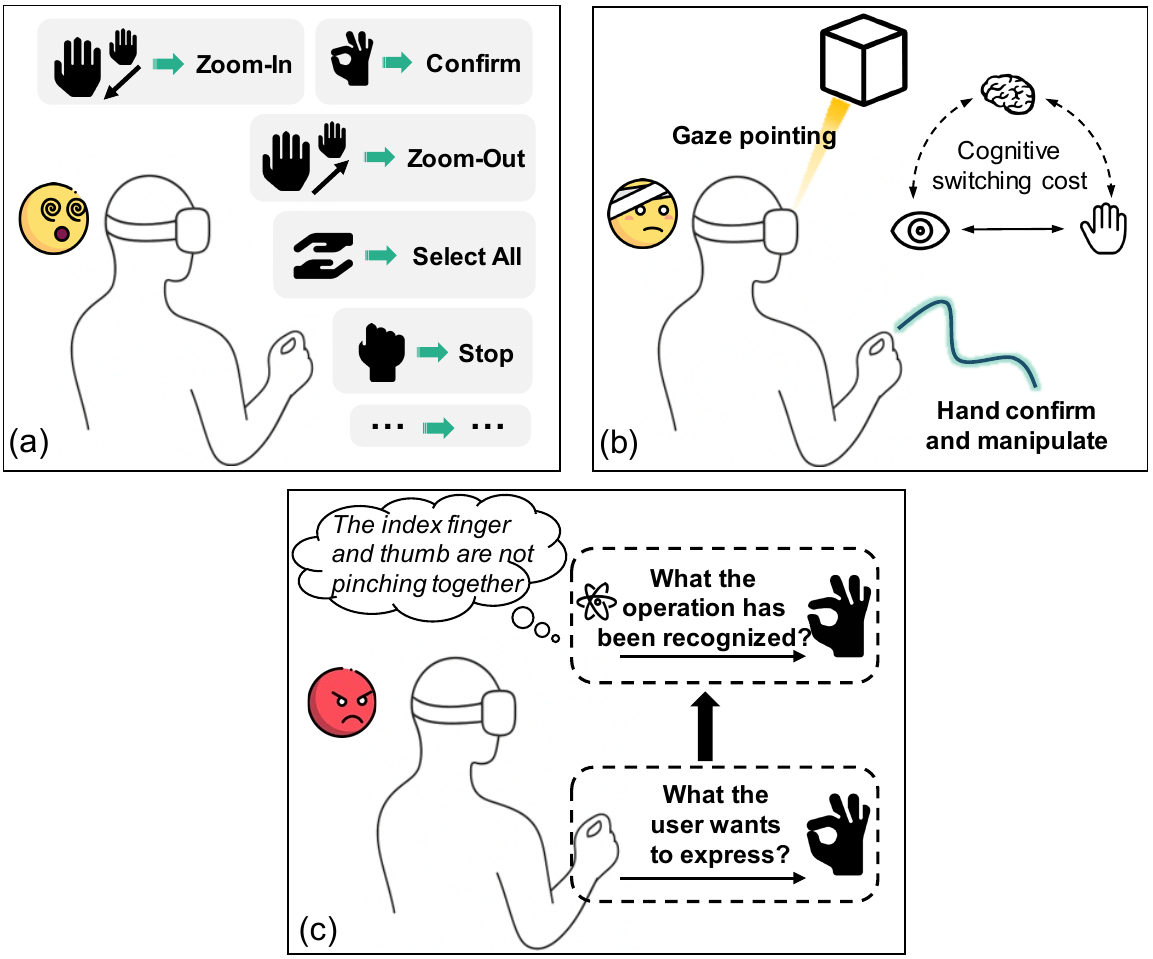}
    \end{center}
    \caption{
       \majorrevise{}{This Operation-to-Intent paradigm presents several limitations. (a) Users must memorize multiple gesture-task associations, raising learning costs. (2) Users need to coordinate gaze and hand movements, increasing complexity and mental burden. (c) Users must achieve precise gesture matching, with recognition accuracy issues boosting interaction errors.}
    }
    \label{fig:limitation}
    \vspace{-5mm}
    \end{figure}

Common interaction paradigms for eye-hand coordination in VR divide the interaction process into three stages \cite{wang2021interaction, DBLP:conf/ismar/ChenGFCL23, Bao2023, yu2021gaze}, as shown on the left of Fig. \ref{fig:teaser}: 1) primary pointing: users direct eye gaze rays to point at a target; 2) confirmation: users utilize pinch gestures to confirm target selection; 3) manipulation: users employ one-hand or two-hand gestures, including grab, drag, rotate, pinch, and tap, in various combinations to complete tasks such as pouring water, writing, \textit{etc}. However, this paradigm adopts an ``Operation-to-Intent" approach that prioritizes low-level operation design over high-level intent. It focuses on mapping predefined gestures to interaction tasks, such as assigning ``come/go/fist" gestures to control image zooming functions (zoom in, zoom out, or stop zooming) \cite{cao2023real}. This approach presents several limitations as shown in Fig. \ref{fig:limitation}. First, users must memorize multiple gestures and associate them with appropriate tasks, significantly increasing learning costs \cite{wang2021interaction}. Second, users need to coordinate attention between gaze positioning and hand gesture execution, which compounds complexity and mental burden. Third, users must perform precise hand gesture matching, and recognition accuracy issues both increase interaction errors \cite{DBLP:conf/chi/PeiCLZ22, DBLP:journals/tvcg/SongDK23}.

To address these limitations, we propose an interaction approach based on an ``Intent-to-Operation" framework, where users first express their intent, the system translates this intent into concrete operations, and then generates an agent to execute the task accordingly. For tasks such as pouring or writing, users express their intent through natural eye-hand movements informed by their own common sense and habits, rather than learning predefined interaction stages or gestures. This design enables intuitive interaction, removing the need for learning or adaptation to specific protocols. By focusing on high-level intent recognition instead of precise gesture matching, our approach accommodates flexible intent expressions and is highly tolerant of recognition errors, as long as the user's actions align with their intents.

Building on this foundation, we leverage Large Language Models (LLMs), which have shown strong capabilities in understanding user intent from text and improving interaction efficiency and usability \cite{chen2024supporting, Yang2025}. \majorrevise{}{Trained on extensive text corpora, LLMs effectively encode common-sense knowledge about human activities \cite{Li2025, chen2023largelanguagemodelsmeet}, enabling comprehension of everyday tasks and associated tools.} However, applying LLMs to VR spatial interaction remains challenging due to the inherent gap between natural language processing and the predominantly spatial, non-linguistic nature of VR interactions.

To bridge this gap, we introduce \textit{\textbf{\ourmethod}}, a spatial interaction agent via LLM-powered eye-hand motion intent understanding in VR, as shown on the right of Fig. \ref{fig:teaser}. \ourmethod~comprises intent recognition and agent-based execution. After users demonstrate their intent through actions, \ourmethod~analyzes eye gaze and hand motion data and translates them into natural language descriptions. The system then employs the LLM to infer user intent and present likely options for user selection. Finally, the LLM generates motion or effect trigger data to guide the agent to execute tasks. We validate our approach through two user studies. The first compares our method with two ``Operation-to-Intent" techniques (controller-based and gaze + pinch (GP) interaction) across over 60 spatial interaction tasks. Our method achieves higher intent recognition accuracy than GP (97.2\% vs 93.1\%) and superior performance across arm fatigue, usability, novelty, and user preference metrics. The second study examines the individual contributions of eye gaze and hand motion channels in intent recognition and explores intent expression via eye gaze combined with speech. Our findings provide valuable insights for advancing intelligent VR interactions.

The specific contributions of our work are three-fold:
\begin{enumerate}
\item We propose a novel ``Intent-to-Operation" interaction paradigm for VR that allows users to express interaction intent through natural eye-hand motions without learning predefined gestures, fundamentally shifting from operation-centric to intent-centric interaction design.

\item We design and implement \textit{\textbf{\ourmethod}}, an LLM-powered spatial interaction agent that translates multimodal eye-hand motion data into natural language descriptions for intent recognition and generates executable agents to complete corresponding tasks.

\item Through extensive user studies across over 60 spatial interaction tasks, we demonstrate that our method achieves higher intent recognition accuracy (97.2\% vs 93.1\%) than gaze + pinch interaction and significantly outperforms traditional ``Operation-to-Intent" techniques in arm fatigue reduction, usability, and user preference.
\end{enumerate}

\vspace{-2mm}

\section{Related Works}
In this section, we review single-modal and multimodal interactions in VR, followed by a discussion of AI and LLM applications in VR.

\vspace{-3mm}

\subsection{Single-Modal Natural Interaction in VR}

Single-modal natural interaction has been extensively researched in VR. We examine three primary modalities: hand gestures, eye gaze, and speech.

\textit{Hand gestures.} 
\majorrevise{}{Valued for their intuitiveness and ability to simulate real-world actions \cite{DBLP:conf/vr/ChaconasH18, GazeRing}, gestures facilitate actions such as pinching, non-linear interaction \cite{go_go_1}, and dual-hand manipulation \cite{DBLP:conf/chi/PeiCLZ22}. However, traditional gesture-based systems require users to memorize predefined gesture vocabularies and perform precise matching \cite{intenselect_2, focalpoint_3}. Extended use leads to muscle fatigue \cite{DBLP:conf/chi/Hincapie-RamosGMI14}, while tracking errors and occlusion  compromise accuracy\cite{focalselect_4, fully_occluded_5}}.

\textit{Eye gaze.} This method is favored for its minimal physical effort and seamless integration \cite{DBLP:journals/tvcg/SidenmarkP0CGWG22, WangTVCG2022}, enabling rapid object selection. However, challenges include the ``Midas touch" problem and reliability issues \cite{DBLP:conf/ismar/MohanGFY18, 10108465}. Gaze-based systems typically require explicit confirmation mechanisms that interrupt natural interaction flow.

\textit{Speech interaction.} Speech enables natural expression through LLM integration \cite{DBLP:conf/vr/GiunchiNGS24} and silent speech recognition offers solutions to social acceptance concerns \cite{DBLP:journals/tvcg/CaiML24}. However, speech interactions face recognition latency, accuracy issues, and increased cognitive load for command memorization \cite{DBLP:conf/vr/SaintAubertAMPAL23}. Traditional speech systems require users to learn specific command vocabularies and syntax.

Single-modal methods fundamentally require users to learn and execute predefined operations to express intentions, highlighting the need for more intelligent approaches.

\vspace{-2mm}

\vspace{-2mm}

\subsection{Multimodal Interaction in VR}

We discuss two multimodal interactions relevant to our work: hand-eye coordination and gaze-speech interaction. 

\textit{Hand-eye coordination}. Recent VR studies leverage gaze for rapid selection and hand gestures for precise manipulation \cite{yu2021gaze, DBLP:conf/chi/KytoEPLB18}. However, this approach typically mandates sequential low-level operations, involving gaze-based selection followed by gesture-based confirmation and manipulation. This requirement increases cognitive load by demanding strict coordination adherence \cite{10108465}.

\textit{Gaze-speech interaction}. This hands-free approach combines gaze-based selection with voice commands for manipulation \cite{DBLP:conf/etra/CastellinaCP08, DBLP:conf/vr/JingLB22}. However, it may hinder precise object manipulation in complex spatial tasks, increasing cognitive load \cite{wang2021interaction, speechdriven}. Speech ambiguity and lack of real-time feedback further limit its utility \cite{GazePointAR}. Users must still decompose their intents into gaze selection followed by appropriate voice commands.

\majorrevise{}{Although recent work has begun incorporating visual attention into multimodal LLMs to refine intent understanding \cite{Rekimoto2025}, the interaction paradigm remains constrained by predefined operation sequences. This dependency imposes high learning costs, highlighting the need for intelligent systems that can interpret user intent from natural multimodal expressions.}

\vspace{-3mm}

\subsection{Applications of AI and LLMs in VR}

AI and LLMs are increasingly integrated into VR systems to enhance intelligence and user experience.

\textit{AI Applications in VR}. AI improves VR by enabling intelligent NPCs \cite{karaca2023ai}, automating virtual environment creation \cite{yang2024amma}, enhancing motion prediction for real-time accuracy \cite{gamage2021so}, and adapting content based on eye movement \cite{10670454}. These applications demonstrate AI's potential to reduce user workload by automating complex behaviors and adapting to user patterns.

\textit{LLM Applications in VR}. 
\majorrevise{}{LLMs represent a paradigm shift in VR, enhancing capabilities ranging from fundamental text entry \cite{chen2024supporting} and spatial interactions \cite{Li2025} to intelligent assistance in creation \cite{Zhu2025} and social contexts \cite{Yang2025}.}
These applications collectively leverage LLMs' superior understanding capabilities to interpret natural language descriptions and generate appropriate virtual content.

LLMs possess extensive common-sense knowledge about human activities, enabling understanding of underlying user intents. By leveraging LLMs' understanding capabilities, VR systems could interpret user intent from natural multimodal expressions, supporting the transition from Operation-to-Intent to Intent-to-Operation paradigms.

\vspace{-2mm}

\section{Motivations and Challenges}

To address the limitations of Operation-to-Intent paradigms, we aim to build an Intent-to-Operation system that enables users to interact naturally based on their own habits and intents. This approach presents several key challenges:

\textit{Q1: How can we enable users to express interaction intent through natural movements without learning predefined gestures?} 
Current Operation-to-Intent paradigms impose high learning costs by requiring users to memorize specific gesture-to-task mappings \cite{Bao2023, wang2021interaction}. This motivates the design of systems that interpret intent from natural hand movements rooted in daily common sense, thereby eliminating the need for explicit gesture training.

\textit{Q2: How can we bridge the gap between spatial VR interactions and language-based AI understanding?} While LLMs excel at understanding user intent from text and possess extensive common-sense knowledge about human activities, VR interactions are predominantly spatial and non-linguistic in nature. How can we effectively convert spatial interaction data into natural language descriptions that preserve essential intent information for LLM processing?

\textit{Q3: How can we achieve robust intent recognition that accommodates flexible intent expressions?} Unlike predefined gesture systems that require precise matching, an Intent-to-Operation approach must handle variations in how different users express the same intent. Users may perform similar tasks with different hand movements, gaze patterns, or interaction styles based on their personal habits and preferences. How can we maintain high tolerance for recognition errors while accurately inferring user intents across diverse motion expressions?

\vspace{-3mm}

\section{Design of \ourmethod}
\label{Smartmethod}

\setlength{\abovecaptionskip}{0pt}
\begin{figure*}
    \begin{center}
    \includegraphics[width=1\linewidth]{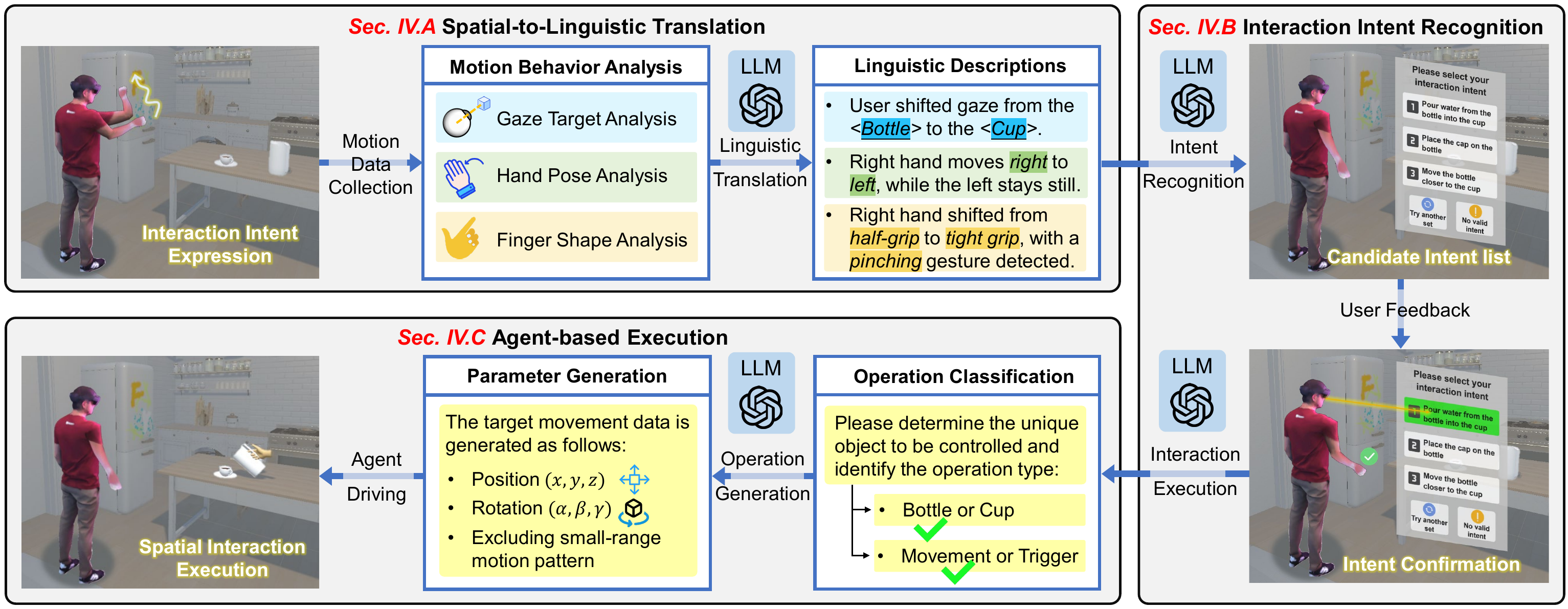}
    \end{center}
    \caption{
    The pipeline of \ourmethod. 
   User eye-hand motions are first captured and translated into natural language descriptions through spatial-to-linguistic translation. The LLM then performs intent recognition to infer possible user intents for selection. Subsequently, the LLM generates executable parameters for agent-driven spatial interaction based on the confirmed intent.
    }
    \vspace{-4mm}
    \label{fig:overview}
\end{figure*}

Building on the Intent-to-Operation paradigm, we design \ourmethod~to include a three-stage pipeline architecture as illustrated in Fig. \ref{fig:overview}. \textbf{Stage 1: Spatial-to-Linguistic Translation} employs LLMs to convert multimodal spatial interaction data (eye gaze and hand motion) into natural language descriptions. This module captures user demonstrations through three complementary channels: gaze target analysis, hand pose analysis, and finger shape analysis, ensuring robustness against individual variations while preserving semantic intent. \textbf{Stage 2: Intent Recognition} leverages LLM's common-sense reasoning to infer user intents from linguistic descriptions and provides ranked intent options for user confirmation. We separate spatial interaction understanding from intent recognition to allow each stage to specialize in distinct cognitive processes, thereby improving overall system accuracy. \textbf{Stage 3: Agent-Based Execution} translates recognized intents into spatial operations, distinguishing between movement operations and trigger operations while using a virtual agent to execute realistic interaction sequences. The detailed implementation of spatial-to-linguistic translation (Section \ref{4.1}), intent recognition (Section \ref{4.2}), and agent-based execution (Section \ref{4.3}) are presented in the following subsections.

\subsection{Spatial-to-Linguistic Translation}
\label{4.1}

For a given task, users first express their intent through demonstrative actions guided by their common sense and habits. To infer user intent, we analyze both visual focus and hand motion data. Visual targets are precomputed by integrating the user's eye gaze with the 3D positions and sizes of scene objects. Hand data analysis encompasses the spatial position and rotation of both hands, as well as gesture configurations, which capture specific hand poses and finger shapes. We utilize LLMs to interpret this multimodal interaction data. The specific analyses are detailed below.

\textbf{1) Gaze Target Analysis}. We identify objects of user interest and attention shifts. \majorrevise{}{Due to the limited accuracy of gaze tracking for VR headsets, our analysis focuses on object-level targeting rather than precise spatial coordinates or specific regions. Based on pilot exploration indicating that users can express their intent within 3 seconds, we set a fixed data acquisition window of 3 seconds.} Eye-tracking data is sampled at 30 Hz. However, we observed that providing all frames to the LLM significantly increases processing time and decreases analysis accuracy due to excessive token length. To address this issue, we downsample the data by selecting every fifth frame, resulting in a total of 18 gaze data points as input. Each data point includes the timestamp, whether the user is fixating on an interactive object, and the target object's name.

\majorrevise{}{The LLM prompt is designed to capture three gaze states: (1) the user continuously gazes at object \textit{A}; (2) the user shifts their gaze from object \textit{A} to object \textit{B}; and (3) the user shifts their gaze from object \textit{A} to no object. This translation is formulated as:}
\begin{equation}
LLM\left(\{T_i\}_{i=1}^{18}, P_{\text{gaze}}\right) \mapsto  D_{\text{gaze}},
\end{equation}
where $T_i$ denotes the gaze target data, $P_{\text{gaze}}$ is the prompt template, and $D_{\text{gaze}}$ represents the natural language description of gaze attention. An example of the LLM input, output, and prompt is shown in Fig. 1(a) of the supplemental material.


\textbf{2) Hand Pose Analysis}.
To infer user intent during object manipulation tasks, it is essential to analyze hand pose changes relative to the user's body coordinate system, as these changes convey rich movement semantics. For example, in a water-pouring from bottle task, the right hand typically moves from right to left relative to the body. Capturing such semantics enables more accurate intent recognition.
We establish the body coordinate system by taking the head position at the start of the task as the coordinate origin. Hand data are sampled at 30~Hz. To reduce data redundancy and processing cost, we compute hand position and rotation angle changes every five frames, resulting in 18 data points per trial.

The LLM prompt is designed to describe hand movement patterns: (1) the movement direction of each hand relative to the body (\textit{e.g.}, left/right, up/down), (2) the magnitude of hand rotation (\textit{e.g.}, whether the rotation is significant), and (3) whether the hands are moving closer to or farther from each other or the head. We formalize the LLM-based translation as:
\begin{equation}
\begin{gathered}
{LLM}\left(\{H^L_i, H^R_i\}_{i=1}^{18}, P_{\text{hand}}\right) \mapsto D_{\text{hand}},
\end{gathered}
\end{equation}
where $H^{L/R}_i$ denotes poses of each hand at frame $i$, including palm position and rotation angle change. $P_{\text{hand}}$ is the prompt template, and $D_{\text{hand}}$ provides a description of hand movement patterns. An example of the LLM input, output, and prompt is shown in Fig.1(b) of the Supplementary Material.

\begin{figure}
    \begin{center}
    \includegraphics[width=0.85\linewidth]{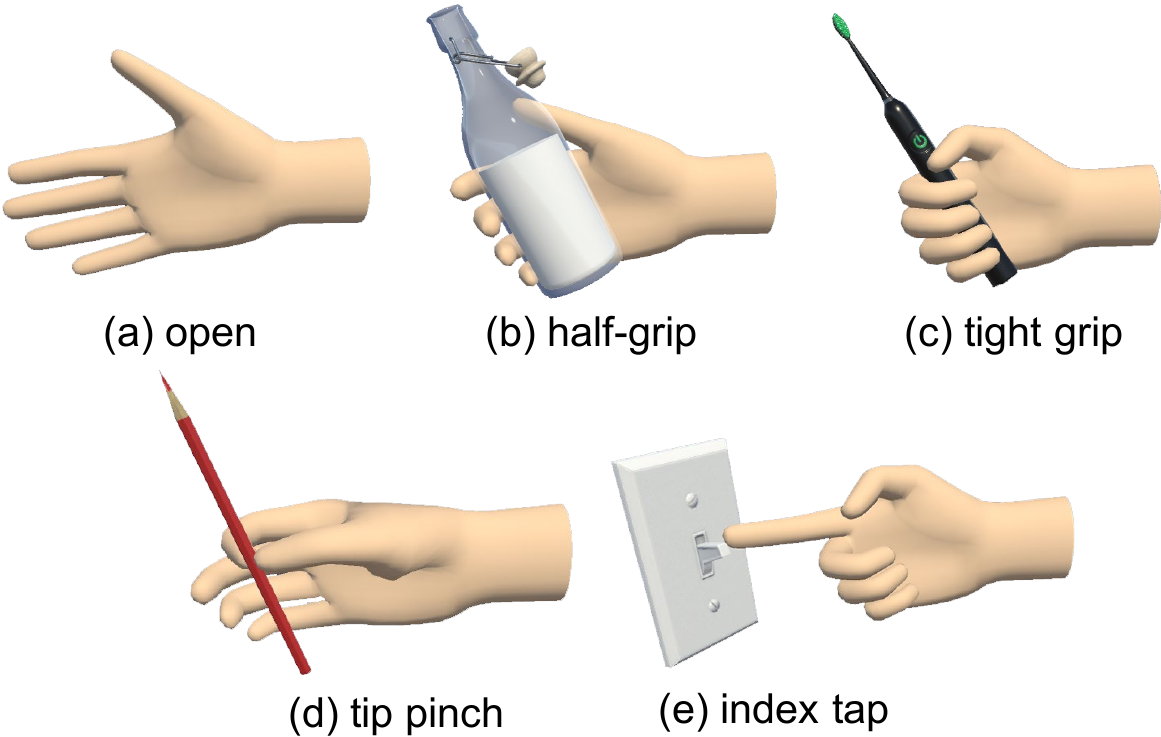}
    \end{center}
    \caption{
       Based on observations of finger shape patterns during hand-object interaction, five hand types are defined: (a) open, (b) half-grip, (c) tight-grip, (d) tip pinch, and (e) index tap.
    }
    \label{fig:finger}
    \vspace{-5mm}
    \end{figure}

\majorrevise{}{\textbf{3) Finger Shape Analysis}. Rather than relying on predefined gestures, we analyze natural interaction patterns to identify five intuitive hand states without requiring user training (Fig. \ref{fig:finger}). We categorize fundamental grasping into open for reaching or releasing, half-grip for partially closing around large objects, and tight-grip for securing items requiring stability \cite{DexCatch2024}. Furthermore, we distinguish fine manipulation states, specifically tip pinch which employs fingertips to grasp slender items, and index tap which extends the finger for activation. This categorization abstracts natural interactions into interpretable states directly tied to user intent.}

\majorrevise{}{{Specifically}, we track dynamic grasp adjustments by quantifying finger flexion at the metacarpophalangeal joint and curl at the distal joints based on established thresholds \cite{Pico}. These continuous values are encoded into Boolean vectors representing the bent or extended status of each finger. The LLM then analyzes these vectors against typical configurations to classify fundamental states (\textit{e.g.}, open, half-grip, tight-grip) and identify fine manipulation gestures like tip pinch or index tap. The translation is formalized as:}
\begin{equation}
\begin{gathered}
LLM\left(\{F^L_i, F^R_i\}_{i=1}^{18}, P_{\text{finger}}\right) \mapsto D_{\text{finger}}, 
\end{gathered}
\end{equation}
where $F^{L/R}_i$ represents vectors of flexion and curl for each finger at frame $i$, $P_{\text{finger}}$ denotes the prompt template, and $D_{\text{finger}}$ provides description of grasp patterns and transitions. An example of LLM input, output, and prompt is shown in Fig.1(c) of the Supplementary Material.

\vspace{-3mm}

\begin{figure}
    \begin{center}
    \includegraphics[width=1\linewidth]{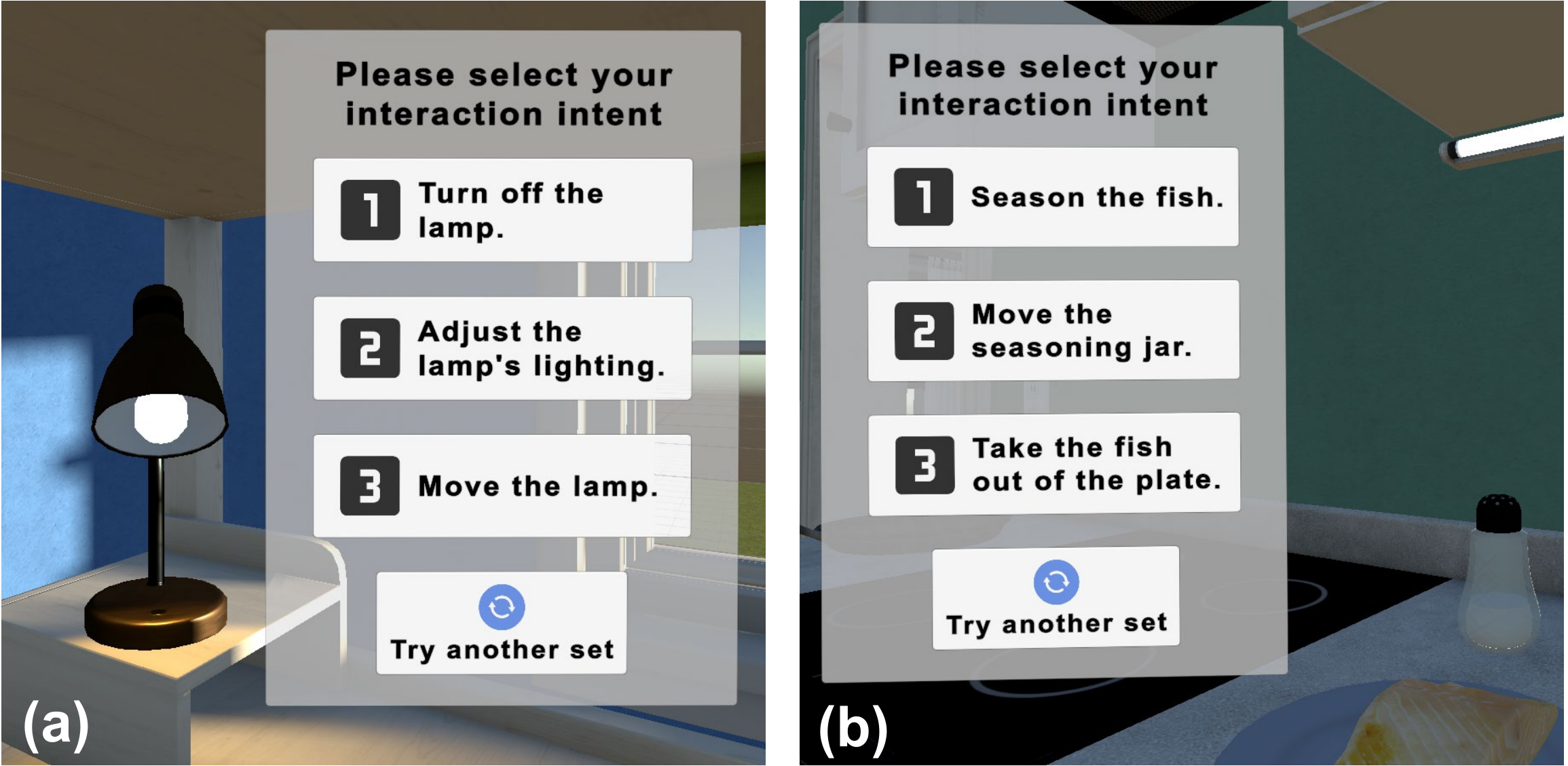}
    \end{center}
    \caption{
        We demonstrate intent recognition results for two tasks: (a) adjusting the lamp's lighting, and (b) seasoning the fish.
    }
    \label{fig:intent}
    \vspace{-5mm}
\end{figure}

\subsection{Interaction Intent Recognition}
\label{4.2}

Based on the linguistic representations derived from eye-hand motions, we infer the user's possible interaction intents. We combine the aforementioned descriptions with contextual object state information to disambiguate user intentions and generate ranked intent predictions. The LLM-based inference is formalized as:
\begin{equation}
LLM(D_{\text{linguistic}}, O_{\text{state}}, P_{\text{intent}}) \mapsto I_{\text{intent}},
\end{equation}
where $D_{\text{linguistic}} = D_{\text{gaze}} + D_{\text{hand}} + D_{\text{finger}}$ represents the concatenated descriptions of eye-hand motions, $O_{\text{state}}$ denotes the current states of target objects, $P_{\text{intent}}$ represents the prompt template for intent recognition, and $I_{\text{intent}}$ provides the ranked list of probable interaction intents with confidence scores.

Target state information $O_{\text{state}}$ captures objects with distinguishable states (\textit{e.g.}, light on/off, door open/closed) to resolve ambiguities where identical gestures may indicate opposing intents. For objects without clear state variations (\textit{e.g.}, a fruit knife), the state is marked as ``no special state."

We design a structured prompt $P_{\text{intent}}$ to guide the LLM's intent recognition process based on three key principles:
\begin{enumerate}
    \item \textit{Multi-object Priority:} When gaze data involves multiple objects, the system prioritizes intents that involve both objects simultaneously and considers interaction sequences. Single-object intents are analyzed only when multi-object interactions are implausible.
    \item \textit{State-based Disambiguation:} The LLM initially analyzes intent without state information to avoid bias. State data is incorporated only when the inferred intent exhibits ambiguity (\textit{e.g.}, open/close actions), serving as a disambiguation factor rather than a primary determinant.
    \item \textit{Ranked Output Generation:} The system generates six ranked intents with probability scores ranging from 0 to 100, with higher scores indicating higher likelihood. The interaction target names in the intents must match those identified in the gaze target analysis.
\end{enumerate}

This design leverages gaze as a natural attention indicator. We maintain semantic understanding as the primary basis while using state information for supplementary disambiguation. The multi-candidate approach accommodates LLM uncertainty and provides alternatives when the primary prediction fails. We highlight high-confidence intents ($\geq$ 90\%) for immediate attention. The prompt structure and example outputs are illustrated in Fig.1(d) of the Supplementary Material.

\textbf{Interactive Intent Confirmation.}
Users select from LLM's ranked predictions, providing explicit feedback for intent determination. The top three intents are displayed initially, with additional options accessible through interface controls, as shown in Fig. \ref{fig:intent}. This interactive process ensures accurate intent capture while generating ground truth data for potential system refinement through learning mechanisms in the future.

\vspace{-2mm}

\subsection{Agent-based Execution}
\label{4.3}

\majorrevise{}{
Once user intent is recognized, we translate it into executable VR operations via a structured LLM-based approach. Our system decomposes spatial interactions into two real-world manipulation patterns: \textit{1) movement operations} that relocate objects in 3D space, such as placing a pencil in a holder, and \textit{2) trigger operations} that activate object-specific behaviors like turning on a lamp. For movement, the LLM generates target position and rotation descriptions to drive object relocation. For triggers, since the LLM cannot directly render visual effects, it selects from predefined object-specific effects based on interaction intent. While current implementation relies on these definitions, recent advances in 3D diffusion models \cite{JointDreamer} offer promising avenues for future intelligent visual generation. The generation process is formalized as:}
\begin{equation}
LLM(I_{\text{intent}}, O_{\text{info}}, E_{\text{effects}}, P_{\text{execution}}) \mapsto M_{\text{move}} \lor T_{\text{trigger}}
\end{equation}
where $I_{\text{intent}}$ represents the recognized intent, $O_{\text{info}}$ contains target attributes (pose, mobility, size), $E_{\text{effects}}$ denotes defined trigger effects for each object, $P_{\text{execution}}$ represents the prompt template for interaction execution, $M_{\text{move}}$ specifies target position and rotation angles for movement operations, and $T_{\text{trigger}}$ indicates selected visual effect for trigger operations.

We design a structured prompt $P_{\text{execution}}$ to guide the LLM's execution generation process based on three key principles, with the example of prompt, input data and output result illustrated in Fig.1(e) of the Supplementary Material:
\begin{enumerate}
    \item \textit{Single-Object Focus:} The LLM identifies a single manipulation target from the intent. Complex multi-step tasks are pre-decomposed into sequential single-object operations (\textit{e.g.}, ``place apple in refrigerator" becomes: open refrigerator $\rightarrow$ move apple inside $\rightarrow$ close refrigerator).
    
    \item \textit{Operation Type Classification:} The system analyzes intent against object capabilities to select between movement operations for spatial displacement and trigger operations for object-specific behaviors. If neither condition is met, no operation is output.

    \begin{figure}
    \begin{center}
    \includegraphics[width=0.9\linewidth]{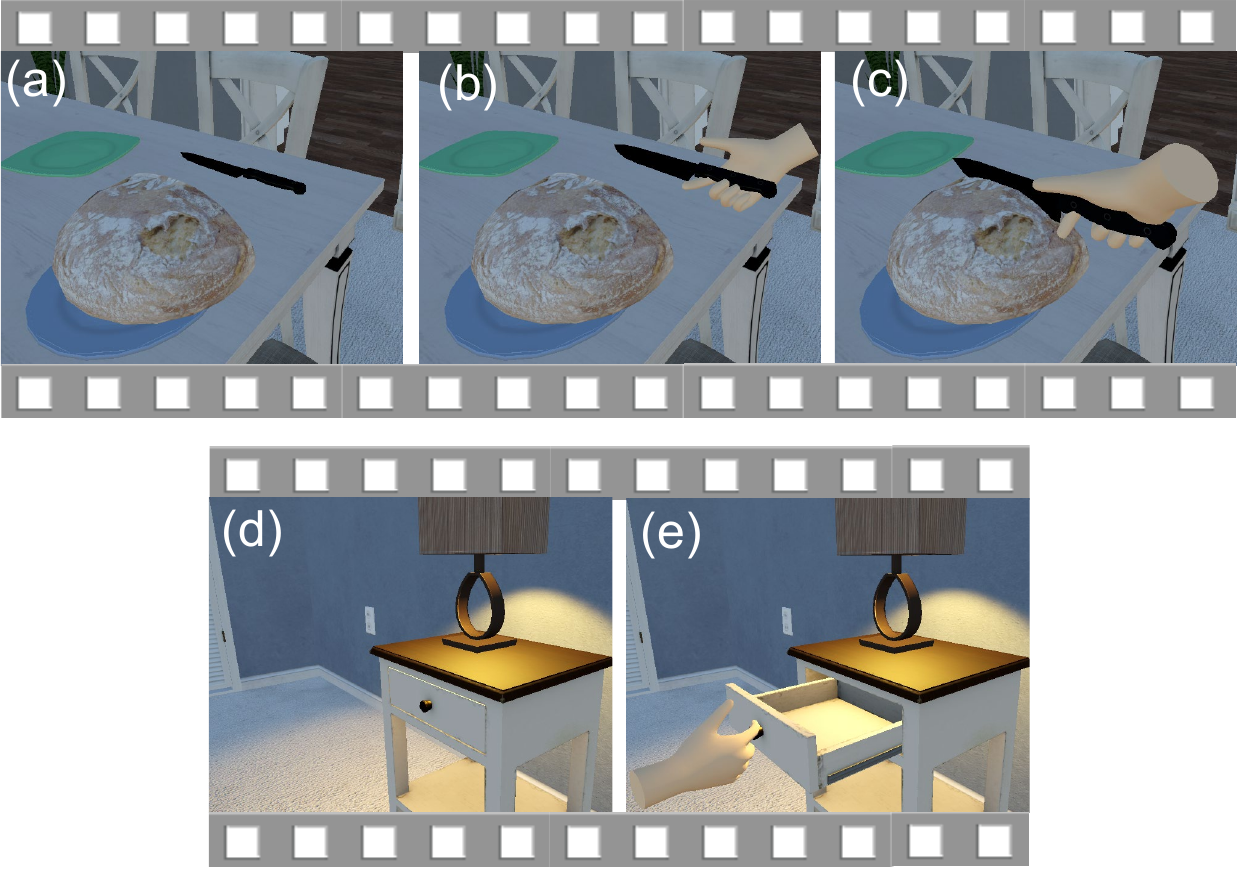}
    \end{center}
    \caption{
       The execution of spatial interactions is based on interaction intents: a) using a knife to cut bread, and b) opening the drawer.
    }
    \label{fig:lantoaction}
    \vspace{-5mm}
    \end{figure}

    \item \textit{Parameter Generation:} For movement, the LLM generates target poses and determines additional small-scale motion patterns are needed to simulate more realistic interactions (\textit{e.g.}, writing with a pencil). For triggers, it selects appropriate visual effects from the predefined set via semantic matching.   
\end{enumerate}

\textbf{Virtual Agent.} 
All LLM-inferred operations are executed by a virtual hand agent (Fig.~\ref{fig:lantoaction}). For movement tasks, the agent grasps and relocates the target to the generated pose; for example, in ``pouring water," the hand agent picks up the bottle and, using the target position provided by the LLM, moves it near the cup and tilts it at the required angle. For trigger tasks, the agent performs contact gestures (\textit{e.g.}, tapping) while the system activates corresponding visual effects, such as illuminating a lamp upon a tap. This approach mirrors real-world manipulation patterns. While we currently use a hand model, recent diffusion models for human-object interaction \cite{wu2024human, jiang2024autonomous} suggest potential for sophisticated humanoid agents in future work.

\vspace{-2mm}

\subsection{System Implementation and Apparatus}

\label{4.4}

Our \ourmethod~is implemented using a Pico 4 Pro VR headset and remote server, with data transmitted via UDP protocol. The headset is equipped with eye tracking, hand tracking, and speech recognition capabilities.
The VR scene was developed using Unity engine (v2022.3.21f1) and rendered at native resolution (2160$\times$2160 pixels per eye) and framerate (90 Hz). The remaining technical steps are executed on the remote server, implemented in Python (v3.10.14). The server calls the LLM API either remotely or locally. The locally deployed LLM runs on an NVIDIA RTX 3090 GPU.

Our prompts employ an instruction-examples structure that accommodates LLMs without relying on version-specific features. This design ensures seamless adaptation to newer LLM versions, such as GPT-4. In the user study, we use ChatGLM-4 (cloud) \cite{glm} unless otherwise specified.

\section{Evalution Design}
\label{5}

\begin{figure}
    \begin{center}
    \includegraphics[width=1\linewidth]{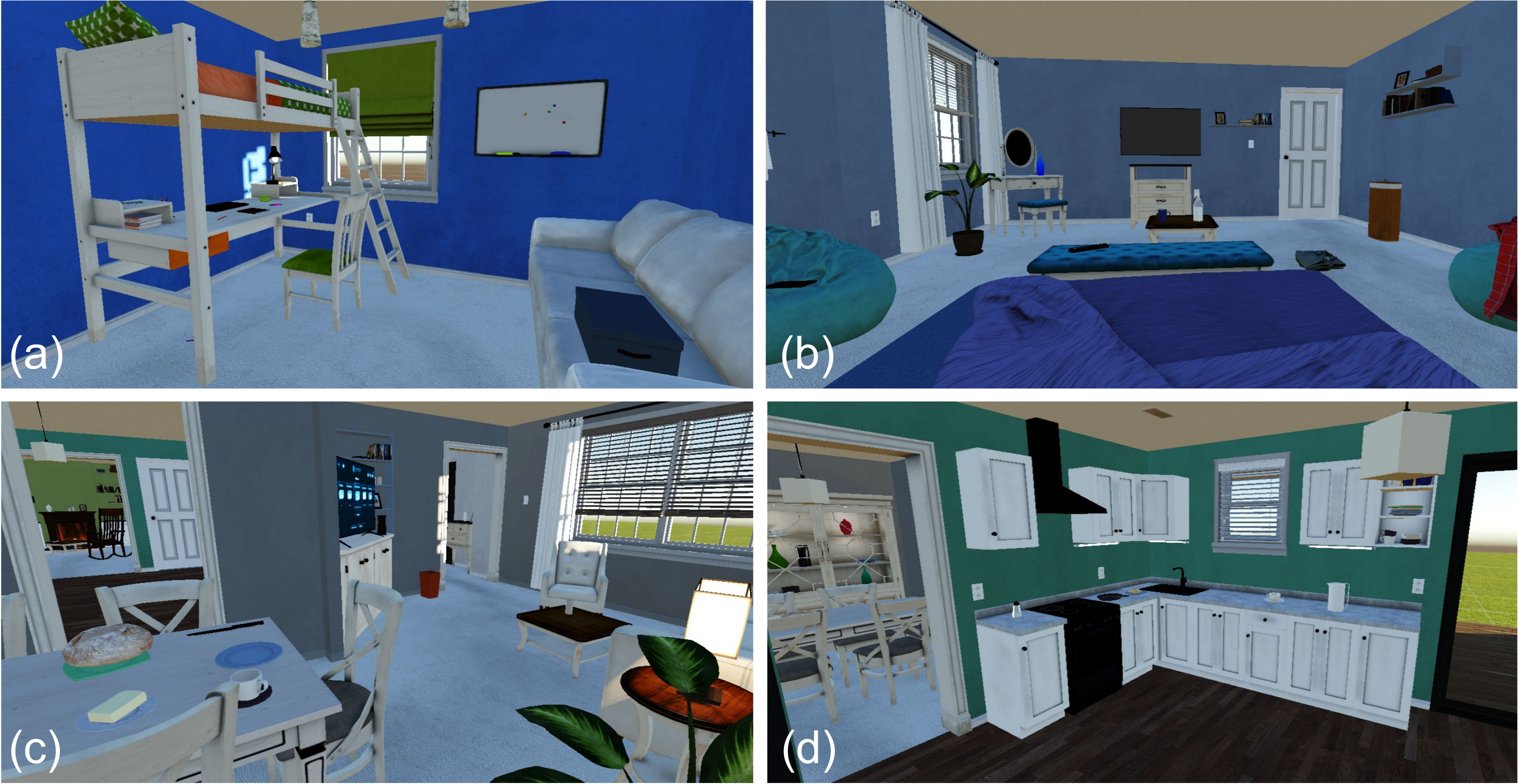}
    \end{center}
    \caption{
       The scenes used in the user study. (a) Study room, (b) Bedroom, (c) Living room, and (d) Kitchen. Scenes (c) and (d) constitute a single, navigable environment in the experiment.
    }
    \label{fig:scenes}
    \vspace{-4mm}
    \end{figure}

\subsection{Research Objectives}

To comprehensively evaluate our \ourmethod~system, we conduct user studies where participants perform typical VR interaction tasks across multiple environments. \majorrevise{}{This evaluation involves both objective and subjective comparisons across different interaction modalities and input channels.} Our objective is to address the following research questions:

\textit{RQ1: How does intent-driven agent execution perform compared to conventional VR interaction modalities?} We investigate whether the Intent-to-Operation paradigm can achieve comparable or superior performance to conventional Operation-to-Intent approaches across objective metrics (time, accuracy) and subjective measures (usability, fatigue, user preference). This question will be investigated in Study 1.

\textit{RQ2: What are the individual contributions of eye gaze and hand gesture channels in intent recognition?} We investigate the effectiveness of combining gaze information with hand gestures compared to using gaze information alone for intent recognition, particularly in ambiguous interaction scenarios where multiple intents are possible. The contribution of each modality channel to the overall system performance will be investigated in Study 2.

\textit{RQ3: How does our LLM-based intent recognition framework perform with alternative input modalities?} To investigate the generalizability of our approach beyond eye-hand motion, we explore the effectiveness of eye-speech input as an alternative modality for intent expression in VR environments. This question will also be investigated in Study 2.

\vspace{-3mm}

\subsection{Participants}

We recruited 16 participants (12 males, 4 females) aged 21-29 years (M = 24.3, SD = 2.2) from our university campus. Pre-study questionnaires using a 5-point Likert scale revealed varying familiarity levels: low with gaze-based interaction (M = 2.5), medium with hand gesture interaction (M = 3.1) and VR systems (M = 3.5), and high with controller-based interaction (M = 3.9). Two participants were left-handed. All participants had normal or corrected-to-normal vision and were fluent in English. The studies were approved by the local ethics committee of Beihang University, and informed consent was obtained from all participants. \majorrevise{}{All participants received a gift valued at approximately 15 CNY (or $\$$2 USD) as compensation for their time.}

\subsection{Task Design and Scenarios}

\majorrevise{}{To address our research questions, we designed 60 interaction tasks within realistic VR environments depicting daily indoor scenarios, as detailed in Tab. I of the Supplementary Material.} These environments provide rich diversity of object types and interaction semantics that users encounter in everyday life, allowing us to evaluate whether our intent-driven approach can handle familiar interaction patterns without requiring users to learn new gestures or protocols. Tasks are categorized into \textit{movement operations} and \textit{trigger operations} as described previously.

\begin{table}[t]
    \centering
    \caption{Twenty-one ambiguous interaction tasks demonstrating multiple possible intents for identical gaze targets.
    }
    \renewcommand\arraystretch{1.4}
    \scalebox{0.92}{
    \begin{tabular}{m{1.6cm}<{\centering}|m{6.9cm}<{\centering}}
    \hline
        \textbf{Objects} & \textbf{Interaction Tasks} \\ \hline
        Laptop & Open it \inblue{}{$\circ$} Close it \inblue{}{$\circ$} Power on it \inblue{}{$\circ$} Power off it \\ \hline
        Desk lamp & Turn on it  \inblue{}{$\circ$}  Adjust its brightness  \inblue{}{$\circ$}  Turn off it \\ \hline
        Window & Open it  \inblue{}{$\circ$} Close it  \inblue{}{$\circ$} Clean it  \\ \hline
        Bottle & Open its cap  \inblue{}{$\circ$} Shake it  \inblue{}{$\circ$} Close its cap \inblue{}{$\circ$} Move it \\ \hline
        Pen and book & Place the pen on the book  \inblue{}{$\circ$} Write on the book with the pen  \\ \hline
        Guitar & Fetch it from a distance \inblue{}{$\circ$} Play it \\ \hline
        Washing machine & Open its lid \inblue{}{$\circ$} Close its lid \inblue{}{$\circ$} Start it \\
        \bottomrule[1.2pt]
    \end{tabular}}
    \vspace{-3mm}
    \label{tab:Tasks}%
\end{table}

To evaluate the contribution of each input channel, we design 21 ambiguous tasks where the same interaction object can support multiple interaction types. These tasks require comprehensive analysis of both gaze targets and hand motion to distinguish user intent. These tasks comprise 11 from the previous task set and 10 newly supplemented scenarios, as detailed in Tab. \ref{tab:Tasks}. The experimental environments are illustrated in Fig.~\ref{fig:scenes} and include the following:

\textbf{Study Room (19 tasks):} This environment provides a study workspace with office supplies (\textit{e.g.}, books, laptop, markers, erasers). This setting allows us to evaluate intent recognition for fine-grained manipulation tasks involving writing, drawing, and document handling where precise hand movements and tool usage patterns are critical.

\textbf{Bedroom (18 tasks):} This environment offers a mix of personal care items and household appliances (\textit{e.g.}, bedside lamp, bottle, washing machine, cups) requiring different interaction strategies. This setting enables evaluation of intent recognition for state-based interactions where the same objects can be opened, closed, turned on, or moved depending on user intent.

\textbf{Living Room \& Kitchen (23 tasks):} This combined environment provides the rich variety of home management activities (\textit{e.g.}, TV cabinet, knife, bread, coffee cup). This setting allows evaluation of intent recognition for complex manipulation sequences involving food preparation, appliance operation, and multi-step cooking tasks that require understanding of procedural interactions.

\vspace{-3mm}

\subsection{Evaluation Metrics}

To comprehensively evaluate our intent-driven agent system, we define metrics that assess task completion performance, system accuracy, and user experience across different interaction modalities.

\textit{Objective Measures}. We define quantitative metrics to evaluate system performance and interaction efficiency.

\begin{enumerate}

    \item \textbf{Task Completion Time:} Mean time to complete tasks. For \ourmethod, we measure time consumed by each component: user intent expression time (U), LLM inference time (L), intent confirmation time (I), and agent execution time (A). We report three time metrics: Total time \textbf{\Agt} = U + L + I + A; total interaction time \textbf{\Agt*} = U + I + A; and user operation time \textbf{\Agt**} = U + I.

    \item \textbf{Task Completion Accuracy:} Average accuracy of task completion. For \ourmethod, we define three accuracy metrics: \textbf{\Agt} (overall accuracy), \textbf{$\Agt^1$} (intent recognition accuracy), and \textbf{$\Agt^2$} (agent-based execution accuracy, assuming 100\% intent recognition accuracy).

    \item \textbf{LLM Performance Comparison:} We evaluate different LLM models (ChatGLM-4 \cite{glm}, ChatGPT-3.5-turbo \cite{gpt}, ChatGPT-4o \cite{gpt}) in terms of inference speed and intent recognition accuracy to identify the optimal model for our framework. \majorrevise{}{These models were selected because they represented the state-of-the-art during the study period (June 2024–January 2025).}
\end{enumerate}

\textit{Subjective Measures}. We assess user experience and acceptance of the Intent-to-Operation paradigm compared to conventional Operation-to-Intent approaches.

\begin{enumerate}
    \item \textbf{NASA-TLX \cite{hart2006nasa}:} We measure cognitive workload including mental demand, physical demand, temporal demand, effort, performance, and frustration using a 21-point Likert scale.

    \item \textbf{Arm Fatigue Level:} \majorrevise{}{We assess physical fatigue using the Borg CR10 scale \cite{borg1998borg} to evaluate whether our approach reduces the physical burden associated with precise gesture execution in traditional paradigms.}

    \item \textbf{Ease of Use:} A 10-point Likert scale measuring learning curve and intuitiveness, specifically whether our method eliminates gesture memorization requirements.
    
    \item \textbf{Novelty:} A 10-point Likert scale assessing users' perception of interaction innovation and naturalness compared to conventional VR interaction paradigms.

    \item \textbf{System Preference:} We collect user preference rankings among the three interaction techniques (\ourmethod, Controller, Gaze+Pinch).
\end{enumerate}

\begin{figure*}
    \begin{center}
    \includegraphics[width=0.95\linewidth]{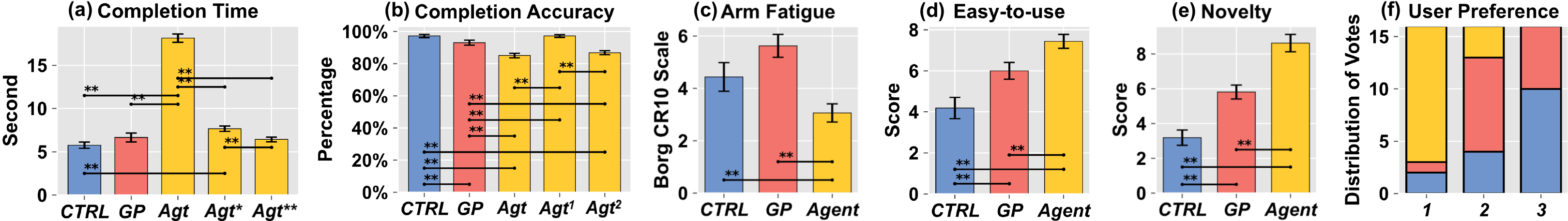}
    \end{center}
    \caption{
       The bar charts display the performance of the techniques across different measurements, with error bars representing the standard error. Statistical significance is indicated by ** for p $\textless$ 0.05. Both \Agt~and \Agent~are our methods.
    }
    \label{fig:result}
    \vspace{-3mm}
\end{figure*}

\vspace{-2mm}

\section{Study 1: Intent-Driven Agent Performance Evaluation}
\label{6}

In this study, our objective is to assess the overall effectiveness of Intent-to-Operation paradigm in VR. We compare our \ourmethod~system with two commonly used VR interaction techniques across multiple interaction scenarios.

\vspace{-2mm}

\subsection{Comparative Setups}

We conduct a within-subjects comparative study evaluating three interaction methods:

\textbf{Controller-based Interaction (\CTRL).} 
We compare controller-based object interaction, extensively explored in object manipulation \cite{hou2021comparison, song2024exploring}.
This interaction consists of two phases: selection and manipulation. \textbf{In the Selection Phase}, users point to the target object using a controller-emitted ray and select it by pressing the trigger button, displaying an operation menu with movement and trigger options. \textbf{In the Manipulation Phase}, for movement tasks, users press the ``A button" to select the movement option and drag the object to the desired location by moving the controller. For trigger operations, users press the ``B button" to display available trigger effects, scroll through options using the joystick, and press the ``A button" to confirm selection, triggering the corresponding visual feedback.

\textbf{Gaze+Pinch Interaction (\textit{GP}).} 
This interaction incorporates the three stages mentioned previously. \textbf{In the Pointing Phase}, users locate the target object by looking at it. 
\textbf{In the Confirmation Phase}, for movement tasks, users pinch their fingers for more than 0.5 seconds to select the object. For trigger operations, users perform two quick successive pinches, displaying available trigger effects. \textbf{In the Manipulation Phase}, for movement tasks, users maintain the pinch gesture while the object follows their hand movements. For trigger tasks, users select the desired effect by gazing at the item and using the pinch gesture to trigger effect selection. To reduce learning costs, we present tasks as menus rather than implementing complex gesture-task associations from previous studies \cite{cao2023real, DBLP:conf/vr/ChaconasH18}, though this approach sacrifices the semantic expressiveness of diverse gesture patterns.

\textbf{SIAgent (\textit{Agent}):} Our method allows users to express interaction intent through natural eye-hand motions. The system recognizes these intents, translates them into concrete operations, and generates agents to execute tasks accordingly.

\vspace{-2mm}

\subsection{Procedure}

\textbf{Preparation Phase:} Participants completed a pre-study questionnaire assessing their familiarity with VR systems and interaction techniques, then signed informed consent. After watching an introductory video explaining the interaction methods and experimental procedure, participants wore the VR headset and completed gaze calibration. To evaluate learning curves, participants practiced two tasks per technique in a selected VR scene for familiarization.

\textbf{Experimental Phase:} Participants performed tasks using all interaction techniques, with technique order counterbalanced via Latin Square design. Scene order was similarly counterbalanced for each technique. Within each scene, participants completed 6 randomly selected tasks. After completing all tasks for a technique, participants took a break and completed an evaluation questionnaire for that technique. Following completion of all techniques, participants ranked their method preferences in a post-study questionnaire. Each participant completed 54 trials total (3 techniques $\times$ 3 scenes $\times$ 6 tasks), with the experiment lasting approximately 70 minutes.

\vspace{-3mm}

\subsection{Results}

We performed repeated-measures ANOVAs ($\alpha$ = 0.05) and post hoc pairwise t-tests to assess whether there were statistically significant differences in a given metric across the various techniques. Prior to using ANOVA, Shapiro-Wilk (S-W) tests and homogeneity of variance tests were conducted.  S-W tests indicated that some data met normality assumptions, while for data that did not satisfy normality, visual inspection of normality test histograms suggested acceptable approximation to normal distribution. Therefore, parametric tests were employed. For data violating homogeneity of variance assumptions, Welch's ANOVA was used.

\majorrevise{}{\textbf{Completion Time.} The results are presented in Fig. \ref{fig:result} (a). Statistical analysis revealed a significant main effect of the interaction method on completion time ($F$(4, 60) = 130.442, $p$ $\textless$ 0.001, $\eta^{2}=0.901$). Specifically, \Agt~was significantly slower than both \CTRL~and \GP~(both $p$ $\textless$ 0.001). This delay is primarily attributed to the three stages of LLM inference, which currently require substantial time. When LLM inference time is excluded, \Agt* remained slower than \CTRL~(7.7 s vs 5.8 s, $p = 0.001$) but showed no significant difference compared to \GP~(6.7 s, $p = 0.075$). Regarding active user interaction time, \Agt** was significantly faster than \Agt* (6.4 s vs 7.7 s, $p = 0.031$) and showed no significant difference compared to the other methods. We have included a supplementary study regarding the impact of latency in the Section I of the Supplementary Material.}

\textbf{Completion Accuracy.} Results are presented in Fig. \ref{fig:result} (b). Statistical analysis revealed that the effect of different techniques and method stages on accuracy was statistically significant ($F$(4, 60) = 22.99, $p$ $\textless$ 0.001, $\eta^{2}=0.529$). Specifically, \Agt~had significantly lower accuracy compared to \CTRL~and \GP~(both $p$ $\textless$ 0.001). Further analysis showed that $\Agt^1$ intent recognition accuracy was not significantly different from \CTRL~(97.2\% vs 97.2\%), and was significantly higher than \GP~(97.2\% vs 93.1\%, $p = 0.02$). However, for $\Agt^2$, which considers agent-based execution accuracy, there was a significant decrease (86.8\%). 
Analysis revealed that accuracy reduction was primarily caused by system timeouts exceeding the 30-second threshold, which automatically terminated agent execution. Our method requires three sequential LLM calls to complete each interaction, with an average API response time of 9.1s (see Tab. \ref{tab:LLM_intent}). During peak usage periods of the GLM-4 cloud service, response times increase substantially, often causing total interaction time to exceed the 30-second limit. We anticipate this limitation will be resolved as LLM computational efficiency continues to improve.

\textbf{LLM Inference Time and Accuracy.}
\majorrevise{}{We recruited three participants from the main study to evaluate intent recognition performance across different LLMs.} As shown in Tab. \ref{tab:LLM_intent}, ChatGPT-3.5-turbo achieved the shortest API call time, followed by ChatGLM-4 (local). ChatGLM-4 (cloud) demonstrated the highest accuracy, with 83.3\% success rate for returning the correct intent on the first attempt and 96.3\% within the top-6 results. ChatGPT-4o (cloud) also exhibited strong performance. \majorrevise{}{Given that API call times will improve as LLMs continue to evolve, we selected ChatGLM-4 (cloud) as our foundational model based on its superior accuracy.}

\begin{figure}
    \begin{center}
    \includegraphics[width=0.95\linewidth]{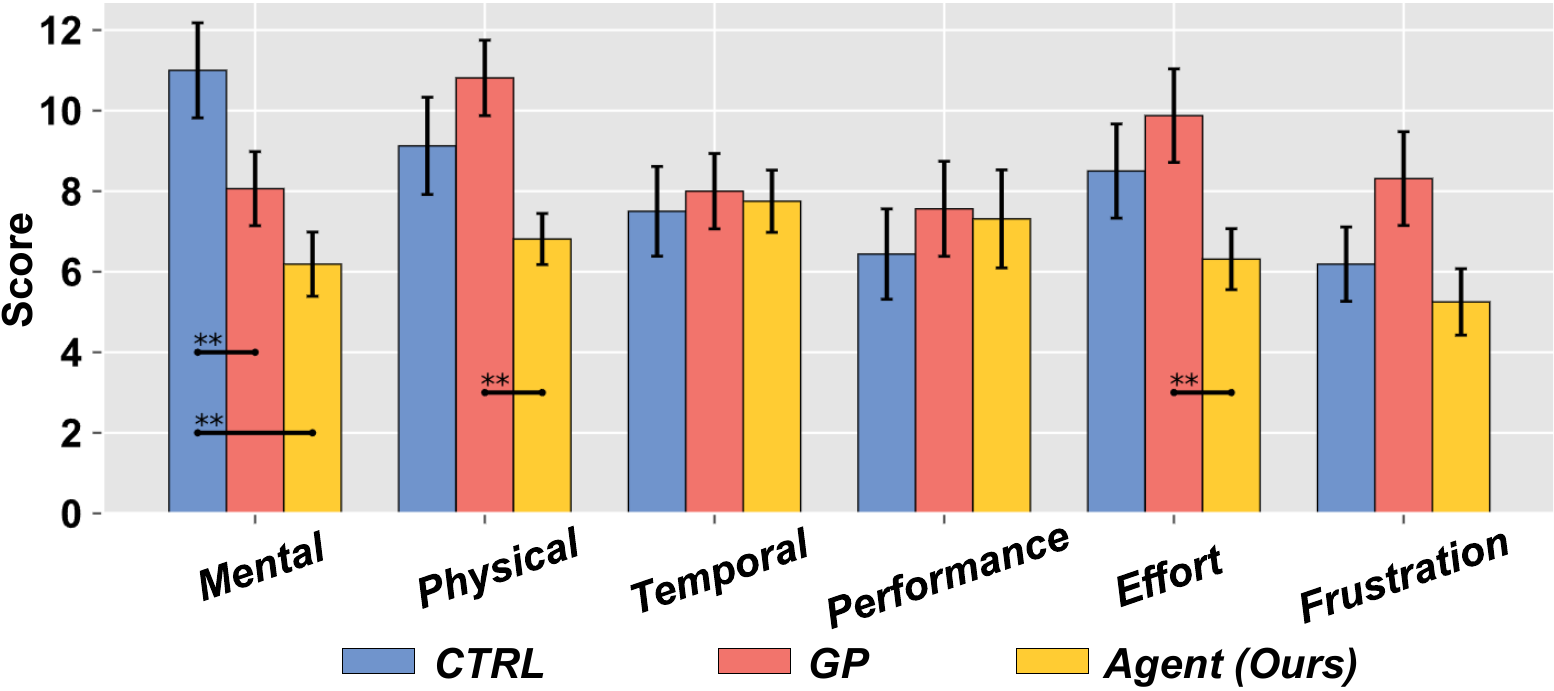}
    \end{center}
    \caption{The bar charts display the NASA-TLX questionnaire scores for the three interaction techniques, with error bars representing the standard error. Statistically significant differences are indicated by ** (p $ < $ 0.05). 
    }
    \vspace{-5mm}
    \label{fig:nasa-tlx}

    \end{figure}

\textbf{Task Load.} Repeated-measures ANOVA on the NASA-TLX questionnaire revealed significant differences in \textit{mental/physical demand and effort} among techniques. 
Post-hoc pairwise t-tests are shown in Fig. \ref{fig:nasa-tlx}. Results showed that \CTRL~had significantly higher \textit{mental demand} than \GP~and \Agent, while \GP~had significantly higher \textit{physical demand and effort} than \Agent~(all $p$ $\textless$ 0.05). \CTRL's higher mental demand stems from cognitive overhead of navigating menus and remembering button mappings, while \GP's elevated physical demand is attributed to sustained pinch gestures and precise hand positioning. Our Intent-to-Operation paradigm reduces both burdens by allowing natural actions without learning.

\textbf{Subjective Experience Metrics.} Fig. \ref{fig:result} (c)-(e) presents results for arm fatigue, ease of use, and novelty assessments, all showing significant differences (all $p$ $\textless$ 0.001). \Agent~exhibited lower arm fatigue than both baselines by eliminating requirements for sustained precise gestures. Regarding ease of use, \Agent~outperformed both techniques by leveraging users' existing common-sense knowledge rather than requiring memorization of gesture-operation mappings. \Agent~also achieved higher novelty scores, reflecting positive user perception of the LLM-powered intent recognition approach.

\textbf{User Preference.} Preference ranking questionnaire results indicate that \Agent~is most preferred by users, as illustrated in Fig. \ref{fig:result} (f). Specifically, 81\% of participants ranked \Agent~first, 12.5\% preferred \CTRL, and one participant favored \GP. This overwhelming preference validates the effectiveness of our Intent-to-Operation paradigm in addressing limitations of traditional VR interaction methods.

\vspace{-3mm}

\setlength{\abovecaptionskip}{0pt}
 \begin{table}[t]
 	\renewcommand\arraystretch{1.4}
 	\centering
 	\caption{Comparison of different LLMs' performance, including API Call Time,  First Intent Accuracy, Top-3 Intent Accuracy, Top-6 Intent Accuracy. }
 		\scalebox{0.92}{
 			\begin{tabular}{m{2.6cm}<{\centering}|m{1.2cm}<{\centering}m{1.2cm}<{\centering}m{1.2cm}<{\centering}m{1.2cm}<{\centering}}
 				\hline
 				LLMs & API call time (s) & 1st Intent Acc. (\%) & Top-3 Acc.(\%) & Top-6 Acc. (\%) \\ \hline
 				GLM-4 (cloud) & 9.1s & \textbf{83.3\% } & \textbf{90.7\% } & \textbf{96.3\% } \\ \hline
        GLM-4 (local) & \underline{7.9s} & 53.7\%  & 75.9\%  & 79.6\%  \\ \hline
        GPT-3.5-turbo (cloud) & \textbf{4.2s} & 46.3\%  & 55.6\%  & 61.1\%  \\ \hline
        GPT-4o (cloud) & 8.5s & \underline{72.2\%}  & \underline{87.0\%}  & \underline{94.4\%}  \\ \hline
 			\end{tabular}%
 		}
 		\vspace{-2mm}
 		\label{tab:LLM_intent}%
 \end{table}%

\subsection{Discussion}

Experimental results demonstrate that the Intent-to-Operation paradigm offers superior user experience over traditional Operation-to-Intent approaches. We organize discussion around three advantages that emerged from evaluation:

\textit{Intuitive Interaction with No Learning Required.}
Traditional VR interaction methods impose substantial learning burdens on users. \CTRL~requires memorizing button mappings and menu navigation sequences, while \GP~demands learning specific gesture-operation associations (\textit{e.g.}, double-pinch for trigger operations). Results show that \CTRL~exhibited significantly higher mental demand than \GP~and \Agent. In contrast, \Agent~leverages users' existing common-sense knowledge about object interactions. Users naturally demonstrate pouring motions when intending to pour water, drawing on everyday experience without learning system-specific protocols. As P12 commented, ``This interaction is intuitive and natural, allowing task completion in your usual way without a fixed operational mode." This design directly contributed to \Agent's superior ease-of-use ratings and reduced mental workload.

\begin{figure}
    \begin{center}
    \includegraphics[width=1\linewidth]{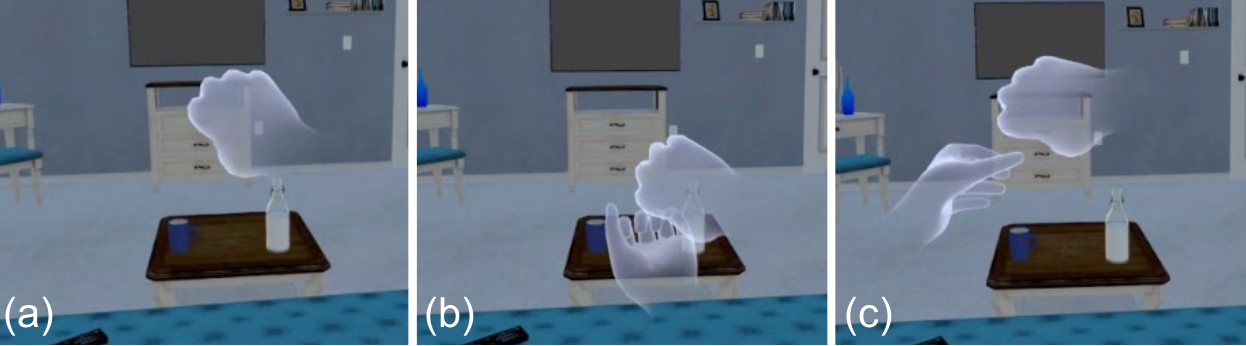}
    \end{center}
    \caption{
       The illustration of flexible intent expression. Users can express milk-pouring intent through various natural gestures: (a) one-hand holding the bottle, (b) two-hand holding the bottle, or (c) using one hand for the cup and one for the bottle.
    }
    \label{fig:expression}
    \vspace{-3mm}
    \end{figure}

\textit{Flexible Intent Expression.}
Our approach accommodates individual differences in motion patterns and interaction preferences. In Fig. \ref{fig:expression}, users express milk-pouring intent through different eye-hand coordination strategies. Traditional systems enforce uniform gesture vocabularies that ignore personal variations. Our LLM-based recognition adapts to diverse expression styles by understanding intent semantics rather than enforcing fixed templates. This flexibility extends beyond hand gestures to different gaze behaviors, movement speeds, and coordination preferences reflecting users' natural habits.

\textit{High Interaction Error Tolerance.}
Traditional VR interaction methods suffer from failure modes where minor execution errors lead to complete task breakdown. \GP~fails when pinch gestures are mistimed or hand positions drift, while \CTRL~requires restarting entire sequences if incorrect buttons are pressed. Our Intent-to-Operation paradigm demonstrates resilience through error recovery mechanisms. When primary intent recognition fails, the system provides ranked alternatives, allowing users to select correct intentions without repeating demonstrations. Additionally, approximate hand motions and imprecise eye gaze convey sufficient information for successful intent inference. P15 noted, ``With \Agent, I barely felt any fatigue during interactions," while P06 and P10 emphasized the system's tolerance of imperfect execution.

The overwhelming user preference (81\% ranked \Agent~first) validates these advantages. P1 remarked, ``\Agent~is innovative, understanding my intent through simple gestures, which is very interesting." Despite current processing speed limitations, users expressed strong confidence in the Intent-to-Operation approach for improving VR interaction experiences.

\vspace{-2mm}

 \section{Study 2: Multimodal Channel Contribution Analysis and Alternative Modality Evaluation}

 \label{7}

To evaluate the individual contributions of eye gaze and hand gesture channels (RQ2) and to explore intent recognition with alternative input modalities (RQ3), we conducted two targeted experiments. The first experiment investigates how hand gesture information improves intent recognition accuracy compared to gaze-only input. The second experiment examines the performance of eye-speech interaction for intent recognition and agent execution, and compares its effectiveness with our eye-hand motion approach.

\vspace{-2mm}

\subsection{Participants and Apparatus}

We recruited 3 participants (3 males) aged 24-29 years from Study 1 for both experiments. This sample size focuses on evaluating technical performance of our LLM-based intent recognition system using objective metrics rather than subjective user preferences, where individual variation has minimal impact on measured recognition accuracy. The apparatus was identical to Study 1, using a Pico 4 Pro headset.

 \begin{table}[t]
 	\renewcommand\arraystretch{1.4}
 	\centering
 	\caption{A comparison of the performance of different LLMs in verifying hand motion for intent recognition, measured by First Intent Accuracy, Top-3 Intent Accuracy, and Top-6 Intent Accuracy. }
 		\scalebox{0.92}{
 			\begin{tabular}{m{2.5cm}<{\centering}|m{2.0cm}<{\centering}m{1.5cm}<{\centering}m{1.5cm}<{\centering}}
 				\hline
 				LLMs  & 1st Intent Acc.  & Top-3 Acc. & Top-6 Acc.  \\ \hline
                Gaze Input Only & 30.2\%	& 63.5\%	& 81.0\% \\ \hline
 				Gaze + Hand Motion &  \textbf{58.3\% } & \textbf{75.0\% } & \textbf{93.3\% } \\ \hline
 			\end{tabular}%
 		}
 		\vspace{-2mm}
 		\label{tab:hand}%
 \end{table}%

\vspace{-2mm}

\subsection{Task Design and Experimental Procedure}

\textbf{Experiment 1: Hand Motion Contribution Verification.} We designed 21 ambiguous tasks where multiple interaction possibilities exist for the same gaze target (\textit{e.g.}, distinguishing between ``open computer," ``close computer," and ``power on computer" when gazing at the same laptop). We compared two conditions: 1) \textit{Gaze Input Only}: LLM receives only eye gaze target sequences without hand motion data; 2) \textit{Gaze + Hand Motion}: LLM receives both gaze targets and complete hand motion information. The experimental procedure followed the same protocol as Study 1.

\textbf{Experiment 2: Eye-Speech Modality Comparison.} We implemented an eye-speech interaction variant using the same tasks from Study 1. Users gaze at target objects and verbally express their interaction intent in natural language. The system performs speech recognition and treats the recognized speech as interaction intent, proceeding directly to agent-based execution as described in Section \ref{4.3}. We compared this modality's performance with eye-hand motion results from Study 1.

\subsection{Results and Discussion}

\textbf{Hand Motion Information Contribution Analysis.} The results presented in Table \ref{tab:hand} demonstrate the critical importance of hand motion information for intent recognition in ambiguous scenarios. When using gaze target information alone, the system achieved 30.2\% first intent accuracy, 63.5\% top-3 accuracy, and 81.0\% top-6 accuracy. However, incorporating hand motion data (including hand pose and finger shape) significantly improved performance to 58.3\% first intent accuracy (+28.1\%), 75.0\% top-3 accuracy (+11.5\%), and 93.3\% top-6 accuracy (+12.3\%). These findings confirm that hand motion information plays a crucial role in disambiguating user intent, particularly for the most likely intent prediction. 

\textbf{Eye-Speech Modality Performance.} Our eye-speech interaction achieved 94.4\% task completion accuracy with an average completion time of 8.4 seconds, demonstrating speech input can accurately express interaction intent. \majorrevise{}{While eye-speech represents a viable alternative, several key advantages favor the hand modality: 1) \textit{Spatial Expression intuitiveness}: Describing complex trajectories (\textit{e.g.}, lift the cup 20 cm while tilting it 45 degrees) imposes high cognitive and verbal effort. Hand motions provide a direct mapping where the input motion mirrors the object's trajectory, avoiding the burden of describing spatial expression; 2) \textit{Fine-Grained Continuous Control}: While speech is well-suited for discrete commands (\textit{e.g.}, ``turn on the lamp''), it is inefficient for smooth, real-time adjustments (e.g., ``make it a bit brighter... a bit more... stop!''). Gestures enable precise, millimeter-level modulation that is cumbersome to achieve through iterative verbal instructions.}

\vspace{-3mm}

\section{Limitations, Future Work and Conclusion}

\subsection{Limitations and Future Work}

\majorrevise{}{Current LLM-based processing introduces significant latency that limits real-time interaction performance.} To address this, future work should explore intent prediction, where LLM inference is initiated based on early gaze and hand cues. Predicting intents in advance allows the system to filter options against ongoing actions, effectively masking latency. Additionally, while our evaluation focused on discrete, single-intent interactions, future research should extend to complex, multi-object sequences.
\majorrevise{}{Furthermore, acknowledging potential LLM hallucinations, we propose integrating traditional gaze-and-hand techniques as a robust fallback for precise control or error recovery.
Finally, our Intent-to-Operation framework naturally aligns with emerging intelligent concepts in future human-AI collaboration, such as cobodied AI paradigms \cite{Lu2026Cobodied}. In particular, AI systems deeply understand human operational intent and execute corresponding actions in real-world environments through robotic platforms, enabling smoother human-robot collaboration in complex physical scenes.}

\vspace{-3mm}

\subsection{Conclusion}

In this work, we introduced SIAgent, an LLM-powered spatial interaction agent that transforms VR interaction from the traditional ``Operation-to-Intent" paradigm to an ``Intent-to-Operation" approach. Our system enables users to express interaction intents through natural demonstrative actions based on their daily experiences and habits, eliminating the need for gesture memorization. The three-stage pipeline effectively translates multimodal spatial data into linguistic descriptions, leverages LLMs for intent recognition, and generates executable interaction agents. Through comprehensive user studies across over 60 interaction tasks, we demonstrated superior intent recognition accuracy compared to conventional gaze-pinch interaction methods, while achieving strong user preference and significantly reduced physical demand. Our multimodal analysis revealed that hand motion information contributes critically to intent disambiguation, substantially improving first intent accuracy. This work advances VR system intelligence through intent-driven design and establishes a foundation for extending such paradigms to human-robot collaboration and embodied AI applications.

\vspace{-3mm}

\section*{Acknowledgments}

This work was supported by the Beijing High Innovation Plan (No. 202504841055), the NSFC (62502020), the Postdoctoral Fellowship Program of CPSF (GZC20252741) and the China Postdoctoral Science Foundation
(2025M774238).


 






\bibliographystyle{IEEEtran}
\bibliography{IEEEabrv,template}












\section{Biography Section}
\vspace{-40pt} 

\begin{IEEEbiography}
[{\includegraphics[width=1in,height=1.25in,clip,keepaspectratio]{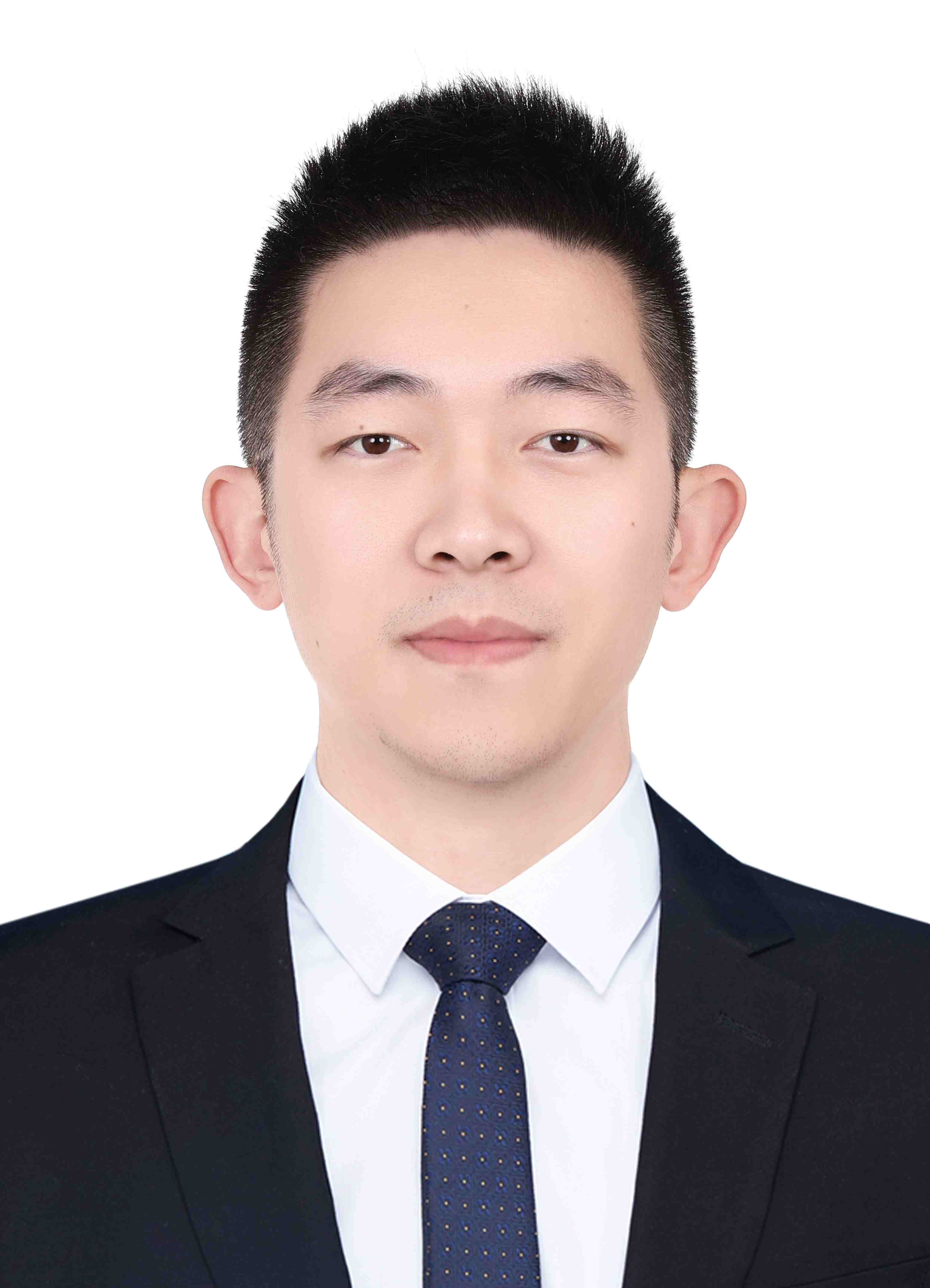}}]{Zhimin Wang} is a postdoctoral researcher with the State Key Laboratory of Virtual Reality Technology and Systems, School of Computer Science and Engineering, Beihang University, China. He received his Ph.D. from Beihang University in 2024 and his B.S. from Chang’an University in 2019. His research focuses on VR/AR, human-computer interaction, and eye-tracking technologies. He serves as a program committee member for AAAI and as a regular reviewer for leading international venues, including IEEE VR, ISMAR, TVCG.
\end{IEEEbiography}

\vspace{-35pt} 

\begin{IEEEbiography}
[{\includegraphics[width=1in,height=1.25in,clip,keepaspectratio]{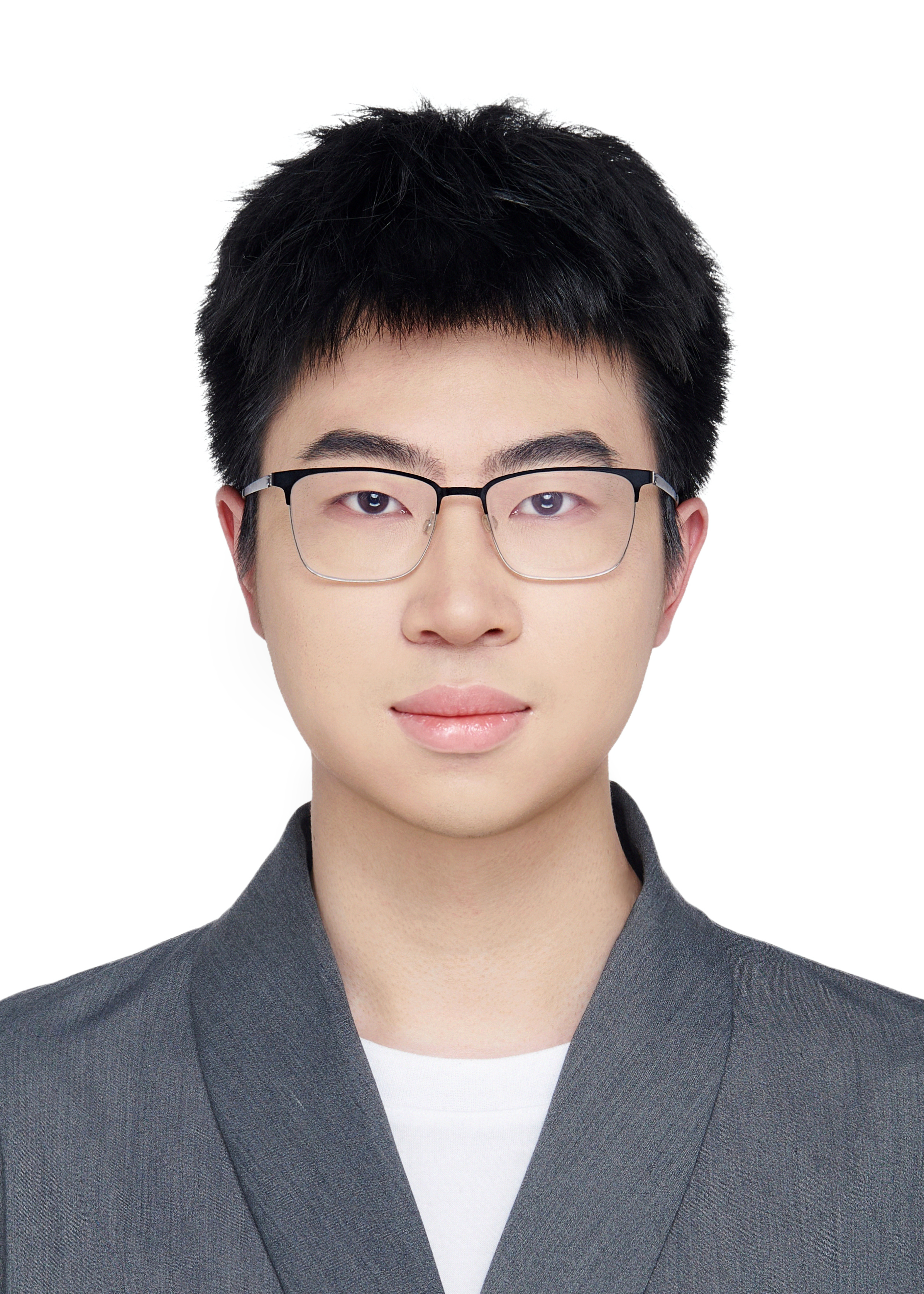}}] {Chenyu Gu} is a Ph.D. student with the State Key Laboratory of Virtual Reality Technology and Systems, Beihang University. He received his bachelor's degree in Computer Science and Technology from Beihang University, China, in 2023. His research focuses on virtual reality, human-computer interaction, and the application of AI in education.
\end{IEEEbiography}

\vspace{-35pt} 

\begin{IEEEbiography}
[{\includegraphics[width=1in,height=1.25in,clip,keepaspectratio]{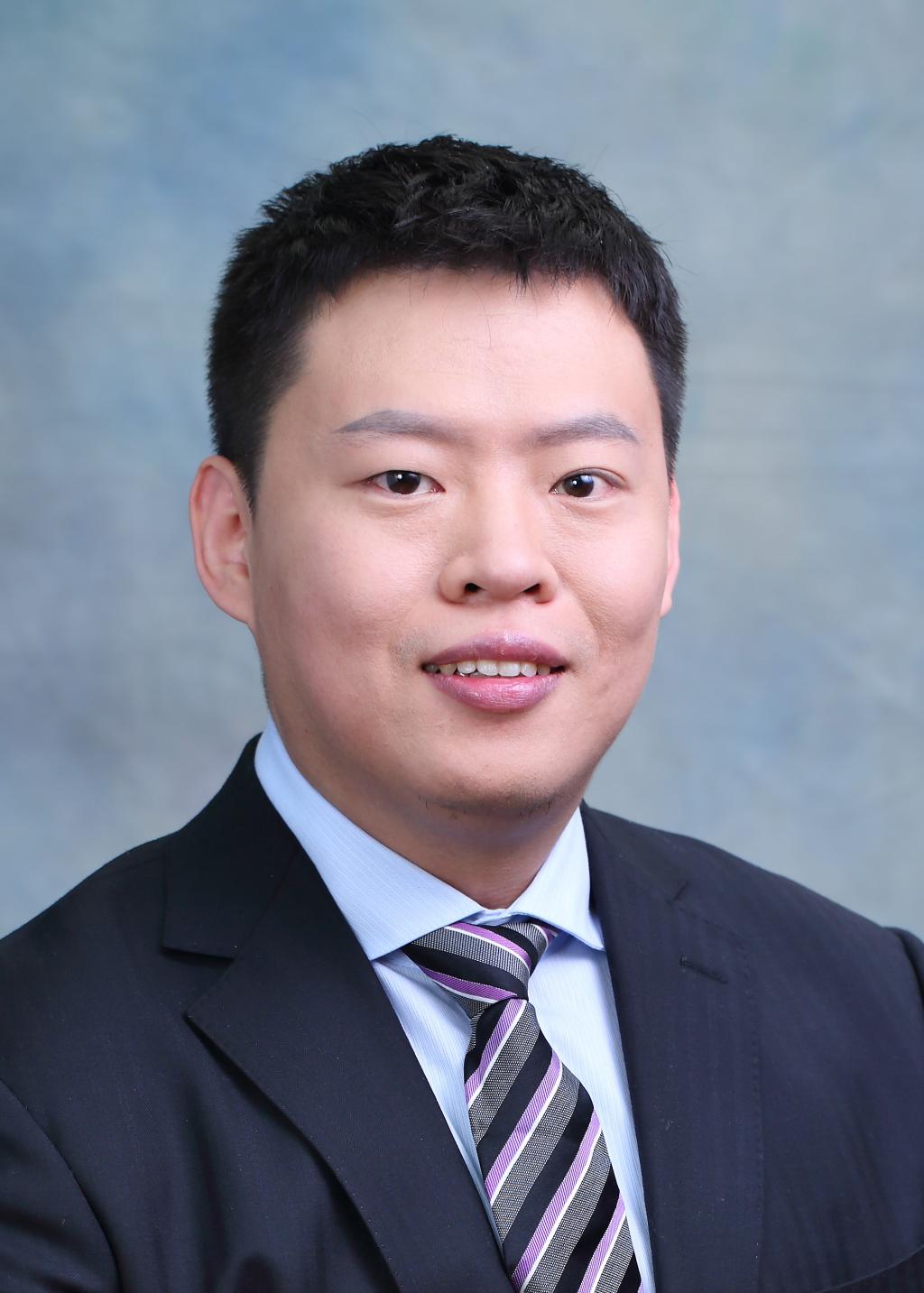}}]{Feng Lu} (Senior Member, IEEE) received the BS and MS degrees in Automation from Tsinghua Univ\textsf{}ersity, China in 2007 and 2010, respectively, and the PhD degree in Information Science and Technology from The University of Tokyo, Japan in 2013. He is currently a full Professor with the State Key Laboratory of Virtual Reality Technology and Systems, School of Computer Science and Engineering, Beihang University. His research interests include computer vision, natural interaction and VR/AR. He is a distinguished member of CCF and CSIG. He is serving/has served as an Area Chair for prestigious international conferences such as CVPR, ICCV, ECCV, NeurIPS and ACM MM.
\end{IEEEbiography}

\vspace{25pt} 

\end{document}